# Quantum physical states for describing photonic assisted chemical changes:
# I. Torsional phenomenon at femtosecond time scale

O. Tapia

Uppsala University, Chemistry-Ångström
Physical Chemistry, Box 523
SE-751 20 Uppsala, Sweden

E-mail: Orlando.tapia@fki.uu.se

Femtosecond torsional relaxation processes experimentally detected and recently reported by Clark *et al.* (Nature Phys. **8**,225 (2012)) are theoretically dissected with a Hilbert/Fock quantum physical (QP) framework incorporating entanglement of photon/matter base states overcoming standard semi-classic vibrational descriptions. The quantum analysis of a generic Z/E (cis/trans) isomerization in abstract QP terms shed light to fundamental roles played by photonic spin and excited electronic singlet coupled to triplet states. It is shown that one-photon activation cannot elicit femtosecond phenomenon, while a two-photon pulse would do. Estimated time scales for the two-photon case indicate the process to lie between a slower than electronic Franck-Condon-like transition yet faster than (semi-classic) vibration relaxation ones.

## 1. Introduction

Infinite in number, quantum states are sustained by finite elementary material constituents; these latter do not permit constructing things in endless sequence, so that object representations of quantum states are not necessarily granted. Besides, there are more reasons suggesting that in abstract Quantum Mechanics (QM) the idea of object (molecule and molecular structure) should recede (though not disappear altogether) and be exchanged by the concept of *quantum state supported by a given materiality*. This latter is a foundational concept [1].

By responding to external probes, the materiality put in evidence the quantum states it withstands. Coherent quantum states are then sustained but not represented as objects by the given materiality; these two extreme models (representation/sustainment) underlie non-commensurate theoretic levels. In this paper we formulate a framework that would help explore this questioning line rooted in our previous works [1-5].

The theoretical scheme developed so far to describe chemical processes [1-5] is readjust to handle torsional degrees of freedom and help analyze grounds for femtosecond torsional phenomena epitomized by Clark's et al. experimental work [6,7]

Yet, so far the word torsional implies semi-classical structural changes, concept not available in *abstract* QM, this situation stays as a non-resolved conflict in the way to understand chemical and physical phenomena, especially when studied with quantum-technologies. Now, embedding quantum base states in quantized photon fields (QPhFs) offer one way to relate abstract Hilbert spaces to relevant laboratory quantum states [1]; in quite specific ways such states may be sensed by real space apparatuses or, as it is the case with double slit devices, can modulate and finally measure the corresponding outgoing physical quantum states [1].

Hypothesis: A fully quantum-physical description of chemical transformations, including correlates for "twisting and turning", must be based on the linear superposition principle built over physical quantum base states; and this must happen beyond standard semi-classical language of potential energy functions and geometric representations for reaction coordinates [8,9]. This paper analyzes some steps in this direction.

## 2. Photon and photonic base states

Photonic base states are sustained by material systems dressed in quantized electromagnetic (qEM) fields; these latter in the guise of photon fields [10]. Matter-sustained base states belong to an electronuclear (EN) class: $\{|E_{j,g(j)}>\}$, see [2]. The photon base states are given in Fock space: $\{|n_\omega>\}$ [11,12]; the label $n_\omega$ indicates number n of energy quanta at frequency $\omega$.

To these complex systems, base states belong to either one of two classes:

1) Non-entangled pairs of photon and matter base states: e.g. direct products, $|E_{j=0,g(0)=0}>\otimes|1_\omega>$,

2) Entangled base states: e.g. $|E_{j=0,g(0)=0};1_\omega>$ no free-photons available yet its presence is implied by labels. Matter I-frame commands.

In general, the frequency ω is a parameter at disposal.

## 2.1 Bare electronuclear quantum states

The complete set $\{|E_{j,g(j)}>\}$ of EN base states [2-5] characterizes all bound electronic states sustained by n-electrons (leptons) and m-nuclei (hadrons) [2,3]. Arbitrary quantum states are written down either as a scalar product (•) or as a linear superposition, albeit in practice both may not be fully equivalent:

$$|\Phi,t> \to (|E_{j=0,g(0)=0}>…|E_{j,g(j)}>…)•$$
$$(C_{j=0,g(0)=0}(\Phi,t)…C_{j,g(j)}(\Phi,t)…)^T =$$
$$\sum_j \sum_{g(j)} C_{j,g(j)}(\Phi,t) |E_{j,g(j)}> \quad (1)$$

Super index $^T$ indicates row vector transposition. The factors entering the scalar product are:

1) An infinite dimensional row vector gathering the basis set $(|E_{j=0,g(0)=0}>…|E_{j,g(j)}>…)$ in an order that time evolution conserves; for the problem at hand no fragments base states (e.g. dissociation, ionization base states) are required; all base states would fulfill the same boundary conditions. 2) The infinite dimensional transposed row vector (column vector) over the field of complex numbers (amplitudes) stand as usual for scalar products: e.g. $C_{j,g(j)}(\Phi,t)=<\Phi,t|E_{j,g(j)}>$; thus a set of non-zero amplitudes characterizes all possibilities the material system might set up to *respond* to external probing at time t; a base state featuring zero amplitude cannot put up a response [1] and, of course, it does not show up in the simple linear superposition form; yet in the column vector it appears as e.g. $0_{j,g(j)}(\Phi,t)$ or $0_{j,g(j)}$.

The linear superposition form is useful for time independent situations with amplitudes confined to finite manifolds. This form conveys, in a sense, truncated information and overlooks possibilities coded by the infinite dimensional vectors.

Thus, the quantum number j characterizes leptonic degrees of freedom; g(j) does it for non-separable EN ones; no classical mechanical models implied only excitation spectra comes in; this is a non-adiabatic – like scheme; Cf.ref. [9]. Thus, any electronic state change is coupled to nuclear degrees of freedom (excitations) and eventually with different specific nuclear states manifolds [2-4]; these latter discernable by spectral responses. For photon-free quantum measurement scheme see ref. [13].

## 2.2 Photon number and photonic base states

Standard Fock base states are used to map free photon number states; energy quantum given by $\hbar\omega = (h/2\pi)(2\pi\nu) = h\nu$; h is Planck's constant. The quantity E=hν *is numerically equivalent to the energy that can be exchanged between EN-systems and radiation field at frequency* ν. This fact, discovered by Planck, initiated the quantum revolution that, with the discovery of lasers prompted a photonic revolution. This amount of energy actually defines the nature of the material *sustaining* the photon base state but not representing materiality as a particle.

A photon state is neither a particle nor a wave in the classical sense; from special relativity theory the energy quantum shows zero rest mass and spin 1; this special EM-material system mediates interactions between quantized EN systems [13]. Only quantum interactions may elicit such physical systems.

## 2.3 Gap-related photonics states

Gap related non-entangled basis state looks like eq.(2):

$$|E_{j=0,g(0)=0}>\otimes|1_\omega> \text{ and } |E_{j=1,g(1)=0}>\otimes|0_\omega> \quad (2)$$

Gap-entangled ones look as:

$$|E_{j=1,g(1)=0};0_\omega> \text{ and } |E_{j=0,g(0)=0};1_\omega> \quad (3)$$

Labels remind the sources (information).

The second term in (2), relates to a radiation-vacuum state $|0_\omega>$ and the matter-sustained excited base state $|E_{j=1,g(1)=0}>$ signal a possible excitation event; the symbol ω labeling for instance $|0_\omega>$ is named stitching frequency.

Non-entangled base states eq. (2) are useful in two ways:
(i) Presenting an incident radiation field state that is being targeted towards an I-frame system [2,3];
(ii) They allow for spontaneous emission in any laboratory direction from a quantum state [1].

For the cases shown, ω is a resonance frequency; this is then a gap-related entanglement.

Caveat: one might be tempted to describe a process of radiation emission as a shift in the material system corresponding pair of basis states, namely:

$$|E_{j=0,g(0)=0};1_\omega> \to |E_{j=0,g(0)=0}>\otimes|1_\omega>. \quad (4)$$

However, this presentation is misleading (if not wrong); it is reviving a particle model again (particle occupying a given EN energy level). A proper description handles quantum states that would involve the *four generic base* (gap related) states from eqs.(2) and (3) taken as a subspace able to sustain coherent states; only amplitudes can change. This new base shares the I-frame of the given materiality [3].

Now, the base vector is organized in two *sectors*; one for the simple direct product and the second sector covers the entangled gap-related base states domain. An entanglement subdomain can be formed with:

$(\ldots |E_{j=1,g(1)=0}\rangle \otimes |0_\omega\rangle \quad |E_{j=0,g(0)=0}\rangle \otimes |1_\omega\rangle \ldots$
$\ldots |E_{j=0,g(0)=0};1_\omega\rangle \quad |E_{j=1,g(1)=0};0_\omega\rangle \ldots)$
$\rightarrow (\;|1\rangle \quad |2\rangle \quad |3\rangle \quad |4\rangle\;)$ (5)

The shorthand notation (5) is used for illustration.

To identify a laboratory state use a *super index* ± in amplitudes vector components, e.g. $(0\;\;1^+\;\;0\;\;0)^T$ that conventionally denotes incoming photonic state, e.g. the (i)-case; conventionally an outgoing state (photonic emission) reads: $(0\;\;1^-\;\;0\;\;0)^T$; either one has infinite number of possible **k**-directions, or in selected cases, for instance, a probing (measuring) device at a definite location [1] would reduce the number of directions. Thus, super- and non-indexed vectors implicitly belong to different spaces yet expressed in the same basis set. The materiality is invariant.

With this notation, a global interaction (emission case (*ii*), Rayleigh scattering) can be read from the following sequence of elementary steps, not processes but *quantum paths* (q-paths):

$(0\;\;1^+\;\;0\;\;0)^T \;{}^+\!\!\Rightarrow (0\;\;1\;\;0\;\;0)^T \rightarrow$
$(0\;\;C_2(t')\;\;C_3(t')\;\;0)^T \rightarrow (C_1(t)\;\;C_2(t)\;\;C_3(t)\;\;C_4(t))^T \rightarrow$
$(0\;\;1\;\;0\;\;0)^T \;{}^-\!\!\Rightarrow (0\;\;1^-\;\;0\;\;0)^T$ (6)

In words:
Laboratory preparation $(0\;\;1^+\;\;0\;\;0)^T$; Transfer to physical Hilbert/Fock-space $(+\Rightarrow)$: $(0\;\;1\;\;0\;\;0)^T$; Time evolution in the entangled manifold $(0\;\;0\;\;C_3(t)\;\;C_4(t))^T$ or extended full space $(C_1(t)\;\;C_2(t)\;\;C_3(t)\;\;C_4(t))^T$; Precursor state $(0\;\;1\;\;0\;\;0)$ mediating transfer to laboratory conditions $(-\Rightarrow)$; Spontaneous emission state: $(0\;\;1^-\;\;0\;\;0)^T$; this is eventually detectable event.

To make operational this generic base states, access to the correct one-photon EM field is required. For eq.(6) one photon enter and another is released. This *entanglement unit* (eq.(5)) plays a central role as one may impose spin and energy conservation. Two-photon case is implemented later on.

Thus, a quantum state $(C_1(t)\;\;C_2(t)\;\;C_3(t)\;\;C_4(t))^T$ would endlessly evolve in time unless special limiting cases are met: e.g. $(0\;\;1\;\;0\;\;0)^T$ that via spontaneous or induced transition leading to $(0\;\;1^-\;\;0\;\;0)^T$ would stop coherence and *a fortiori* entanglements.

Reading the base states necessarily including amplitudes one can get information about the way the materiality (i.e. energy-spin) might be changing. Yet, one is far from any classical physical picture to the extent the amplitudes **cannot** be understood in "population" terms; fully coherent quantum states develop including both sectors; dynamically entangled states cover the full dimension of eq.(5). The standard statistical interpretation of QM [11] does not follow in a natural manner. See refs.[1,12] for more details.

## 3. Z/E Torsion: Photonic assisted Phenomenon

A torsional phenomenon would relate, by definition, two different EN-manifolds herein identified with primed and double primed quantum numbers: j'g(j') and j''g(j''). The process elicits electronic state change [5] at variance with standard chemical perceptions based on semi-classic potential energy functions [8,9].

Then two basic elements are required to start up a quantum theoretical characterization of "torsional" phenomena:

1) Two j-quantum numbers assigned to closed shell state conformers, e.g. Z and E;
2) Subsets of electronuclear quantum numbers including excited j'-, j''-states where angular momentum labels will provide grounds to construct quantum mechanisms.

Constraint 1) implies spectra originated from Z and E states can be characterized independently. For constraint 2), the conversion would always imply base states having j'- and j''- related excited states. This quantum physical requirement is necessary but might not be sufficient to follow up this type of process. Let us briefly elaborate this issue below.

### 3.1 Model Z/E base states

Consider two different base sets:
$Z=\{|E_{j',g(j')}\rangle,\ldots,|E_{j'+1,g(j'+1)}\rangle,\ldots,|E_{j'+2,g(j'+2)}\rangle,\ldots\}$ &
$E=\{|E_{j'',g(j'')}\rangle,\ldots,|E_{j''+1,g(j''+1)}\rangle,\ldots,|E_{j''+2,g(j''+2)}\rangle,\ldots\}$ (7)

Each sequence associated to a particular ground state $|E_{j'=0,g(j')=0}\rangle$ and $|E_{j''=0,g(j'')=0}\rangle$. To mimic double bond features, take nodal planes for Z and E perpendicular [5] (sort of chemical picture); note, same I-frame.

The lowest electronic excited states for j'-family $|E_{j'+1,g(j'+1)}\rangle$ and $|E_{j'+2,g(j'+2)}\rangle$ is modeled as one-electron excitations; $|E_{j'+1,g(j'+1)};{}^1\Pi\rangle$ is spin-singlet and a related spin-triplet $|E_{j'+2,g(j'+2)};{}^3\Delta\rangle$, the space angular momenta labeling is used as indication only of parity properties.

The first excited state belongs to $^1\Sigma \rightarrow {}^1\Pi$ spin singlet excitation; the second $^1\Sigma \rightarrow {}^3\Delta$ is spin triplet excitation, forbidden in first order, where the mixing is due to spin-orbit interaction via $^1\Pi$ base state. A q-path such as $^1\Sigma \rightarrow {}^1\Pi \rightarrow {}^3\Delta$ naturally introduces a *torsional concept as involving an angular momentum change*.

The lowest electronic excited states for j''-family $|E_{j''+1,g(j''+1)}\rangle$ and $|E_{j''+2,g(j''+2)}\rangle$ are also modeled as one-electron excitations; they constitute a spin-singlet, say $|E_{j''+1,g(j''+1)};{}^1\Pi\rangle$ and spin-triplet $|E_{j''+2,g(j''+2)};{}^3\Delta\rangle$ subset.

Note, $\langle E_{j'=0,g(j')} | E_{j''=0,g(j'')}\rangle = 0$ and both ground states show same parity (both electronic closed shells). No direct EM transition is hence allowed. This justifies calling them different "chemical species".

Particular case: ground state energy levels fulfill inequality:
$$E_{j'=0,g(j')=0} > E_{j''=0,g(j'')=0} . \qquad (8)$$
This energy gap plays a key role to be discussed later on. The conventional ground state for E-conformer (j'') is more stable than the Z-conformer (j').

**3.2 Gap-related (g-r) photonic basis sets**
Let us write the relevant g-r-photonic base states for a 1-photon activation process; put non-entangled sector first and thereafter the entangled components, one gets:

$(\ldots|E_{j'=1,g(j')=0}\rangle \otimes |0_{\omega'(01)}\rangle \quad |E_{j'=0,g(j')=0}\rangle \otimes |1_{\omega'(01)}\rangle \ldots$
$|E_{j'=2,g(j')=0}\rangle \otimes |0_{\omega'(02)}\rangle \quad |E_{j'=0,g(j')=0}\rangle \otimes |1_{\omega'(02)}\rangle \ldots$
$|E_{j''=1,g(j'')=0}\rangle \otimes |0_{\omega''(01)}\rangle \quad |E_{j''=0,g(j'')=0}\rangle \otimes |1_{\omega''(01)}\rangle$
$|E_{j''=2,g(j'')=0}\rangle \otimes |0_{\omega''(02)}\rangle \quad |E_{j''=0,g(j'')=0}\rangle \otimes |1_{\omega''(02)}\rangle \ldots$
$|E_{j'=1,g(j')=0};0_{\omega'(01)}\rangle \quad |E_{j'=0,g(j')=0};1_{\omega'(01)}\rangle \ldots$
$|E_{j'=2,g(j')=0};0_{\omega'(02)}\rangle \quad |E_{j'=0,g(j')=0};1_{\omega'(02)}\rangle \ldots$
$|E_{j''=1,g(j'')=0};0_{\omega''(01)}\rangle \quad |E_{j''=0,g(j'')=0};1_{\omega''(01)}\rangle \ldots$
$|E_{j''=2,g(j'')=0};0_{\omega''(02)}\rangle \quad |E_{j''=0,g(j'')=0};1_{\omega''(02)}\rangle \ldots ) \quad (9)$

The gap is indicated as sub index of the photon number label. Introduce a simplified notation: ($|1\rangle \ldots |16\rangle$); 1-to-8 un-entangled and 9-to-16 stand for entangled base state components. Amplitudes indicated as:
$$\mathbf{C} = (C_1 \; C_2 \; \ldots \; C_7 \; C_8 \; C_9 \; C_{10} \; \ldots \; C_{15} \; C_{16})^T \quad (10)$$
The coherent quantum state introduces dynamic correlations including the direct product base sector.

**4. Quantum isomerization: mechanisms**
**4.1. Z/E q-paths**
Here, torsional change can be highlighted in a rough manner. Start up at the activation step:
$$(0_1 \; 1^+_2 \; \ldots 0_6 \; 0_7 \; 0_8 \; 0_9 \; 0_{10} \; \ldots \; 0_{15} \; 0_{16})^T . \quad (10')$$
A "successful" path:
$$(0_1 \; 0_2 \; \ldots 1^-_6 \; 0_7 \; 0_8 \; 0_9 \; 0_{10} \; \ldots \; 0_{15} \; 0_{16})^T \quad (10'')$$

This vector signals a photon emission leaving a base state for the j'' domain. Torsion as a state change must somehow be achieved.

How would the physics be in order to attain such transformation? Note that energy conservation forbids this path; parity stays in the direct way also.

But there are other elements deserving attention. A first clue suggests that to produce such a change of pattern, from (10') to (10'') it is necessary to enter the Hilbert-Fock space including both sectors first: i.e. using *external* couplings quantum time evolution would be enforced. Thus, amplitudes propagates from $(0_1 \; 1_2 \; \ldots 0_6 \; 0_7 \; 0_8 \; 0_9 \; 0_{10} \; \ldots \; 0_{15} \; 0_{16})^T$ i.e. the expanded Hilbert/Fock space $\mathbf{C}(t)$. A q-state found in a vicinity of $(0_1 \; 0_2 \; \ldots 1_6 \; 0_7 \; 0_8 \; 0_9 \; 0_{10} \; \ldots \; 0_{15} \; 0_{16})^T$ could show decoherence via state (10''). The event itself can't be predicted in time but it is a possibility.

Yet, an indirect process would do if internal energy is channeled in supplement to the incoming photonic field, $\hbar\omega'(01)$; the energy sought is associated to the g(j'')-domain; but now it is related to non-radiative transformations, Cf. ref.[7,9]. In convokes a different complexity level (not analyzed here).

If only non-radiative processes mediate the transformation then the lifetimes will be way beyond the femtosecond range and probably be found deep within the slower range characteristic of the vibration relaxation model [7].

What about the femtosecond range then? Obviously it cannot be accessible with only one-photon input.

**4.2 Z/E two-photon process**
Thus there is need for a second independent photonic source to open reactive channels. One method would be to shine energy at $\omega''_{(01)}$ in the E-channel but in order to be effective the amplitude must be different from zero; however, by hypothesis it is zero as long as channel j'' is physically closed; a reason for this is that by looking at the label for $|E_{j''=0,g(j'')=0};1_{\omega''(01)}\rangle$ the system has no energy equivalent to $1_{\omega''(01)}$ and the amplitude must be zero to fulfill energy conservation. While, the energy level for $|E_{j''=1,g(j'')=0};0_{\omega''(01)}\rangle$ enters quantum interactions and amplitudes can be included in the coherent state domain.

The femtosecond experiment reported by Clark et al.[6,7] contains a sequential two-photon process. Two photon states (two spin units) can be recombined to get relevant high frequencies and appropriate energy gaps. And to accomplish the feat the materiality of the material system is required (not objects). In one word, the one-photon state $|1_{\omega''(01)}\rangle$ amplitude must be brought up to non-zero values.

Consider first a mechanistic gist of the quantum conformation change:

The base $|E_{j'=1,g(j')=0};0_{\omega'(01)}\rangle$ maps onto $S_1$ [6] that in the experimental setup is obtained with a visible-light pump. Here, the push frequency is assigned to $|1_{\omega'''(1n)}\rangle$ the stitch frequency relates $S_1$ to $S_n$ (notation from [6]). The non-entangled input excited base should read:

$|E_{j'=1,g(j')=0};0_{\omega'(01)}\rangle \otimes |1_{\omega'''(1n)}\rangle$.

The base state with two stitching frequencies, that is

$|17\rangle = |E_{j'=n,g(j')=0};0_{\omega'(01)},0_{\omega'''(1n)}\rangle$ is required.

This base state belongs to a full-entangled sector; this expanded space is imposed by the experimental setup; theoretically all pairs of stitching frequencies must be included. Of interest to us is that these new base states would help understand how to develop amplitudes at the j''-channel. One reason for success is that now there is EM energy enough to pay for the excitation/de-excitation gap $1_{\omega''(01)}$ associated to the $^1\Sigma \leftarrow ^1\Pi$ spin singlet at j''-channel.

Experimentally, energy associated to $|17\rangle$ is above $|13\rangle$ and amplitude at $|E_{j''=1,g(j'')};0_{\omega''(01)}\rangle$ could now develop. Otherwise a non-radiative path is still possible but it is of no relevance or interest. The entanglement unit at j''-channel would become active now:

$(\ldots |E_{j''=1,g(1)=0}\rangle \otimes |0_{\omega''}\rangle \quad |E_{j''=0,g(0)=0}\rangle \otimes |1_{\omega''}\rangle \ldots$
$\ldots |E_{j''=0,g(0)=0};1_{\omega''}\rangle \quad |E_{j''=1,g(1)=0};0_{\omega''}\rangle \ldots)$
$\rightarrow (\ |1''\rangle \ |2''\rangle \ |3''\rangle \ |4''\rangle \ )$  (5'')

The difficulty found to get a simple description lies in the change of energy scale for the E and Z isomers. Once the gap is included one understand that Z-gaps and E-gaps differ in a constant that is given in eq.(8). Energy and angular momentum conservation enter the scene. And, even if resonance condition are met between energy levels j'g(j') and j''g(j'') to prompt photon emission at E-channel starting from Z-channel not only the accessibility of gap-determined entangled base states, at a laboratory level, but also must fulfill energy and parity conservation laws.

The one photon case with frequency adequate to pay the Z-E-gap no mechanism is available to activate the j'-to-j'' process. While for a sequential one, the second photon may couple j'-j'' transition, e.g. $^3\Delta_{j'} \rightarrow\ ^1\Pi_{j''}$: this would be enough to credit for the opening.

### 4.3. Where is the torsional phenomenon?

A conformational change turns out to be describable in terms of subtle quantum physical displacement of amplitudes involving q-steps e.g.:

First photon : $^1\Sigma_{j'} \rightarrow ^1\Pi_{j'} \rightarrow ^3\Delta_{j'}$
Second photon : $^3\Delta_{j'} \rightarrow ^1\Pi_{j''} \rightarrow ^1\Sigma_{j''}$
or/and

$^1\Pi_{j'} \rightarrow ^3\Delta_{j''} \rightarrow ^1\Pi_{j''} \rightarrow ^1\Sigma_{j''}$  (11)

Torsion reflects the electronuclear quantum process by variation of angular momenta. Moreover, the j'→j'' q-step associate an orientation change of the transition dipole. The -⇒ process would lead to one photon state emission, $\hbar\omega''_{(01)}$ carrying out a spin 1. Therefore, the two units of spin are used differently: one changes j' angular momentum, the other while changing angular momentum at j'' would "evacuate" the excess spin with a photon state. This accomplishes the quantum physical torsion.

What about life times? A qualitative answer follows from the preceding analysis. The cascade reorganization (of quantum numbers) may have the character of electronuclear rearrangement where the singlet-triplet subspace mediate time evolution until getting at resonance zone with the outgoing channel that corresponds to a j'→j'' transition in the entangled unit eq.(5''). Being an electronic (or better electronuclear) transition it may be relatively fast compared to pure vibration mediated process.

### 5. Quantum evolution

The base set formed by (5) and (5') permits a q-process starting with the initial state eq.(10'). Because QM ought to include all possibilities the initial state looks like:

$(0_1 \ 1_2(t_o) \ \ldots 0_6 \ 0_7 \ 0_8 \ 0_9 \ 0_{10} \ \ldots \ 0_{15} \ 0_{16})^T$  (10''')

This vector belongs to abstract Hilbert/Fock space so that Schrödinger time evolution may start at initial time $t_o$. The model refers to a closed space where $\mathbf{C}(t)$ evolves under the spell of Schrödinger equation :

$i\hbar \partial \mathbf{C}(t)/\partial t = (H+V) \mathbf{C}(t)$  (12)

H is the Hamiltonian constructed with the basis set (9) and V includes operators allowing for state mixing. Eq.(12) drives then a coherent evolution in $\mathbf{C}(t)$.

At a non-predictable time say t* the system may show signs of decoherence when for instance one senses the state vector $\mathbf{C}(t^*)$ transition to a neighborhood of state (10''). The fully isomerized state can be sensed and distilled from the quantum system.

Now, the global change might well be found in the femto second range, namely, slower than a pure electronic transition yet faster that vibration-like relaxation.

Again, all these amplitude changes are sustained by the same materiality; classical mechanical models do not help understanding the process. The theory requires non-zero amplitudes to respond to external probes [1].

### 6. Pictorial descriptions

Yet, some pictorial-like presentation may help sensing quantum aspects improving communication channels as well as publicity.

### 6.1. From laser experiments

Clark et al.[6] reported unusual timescales suggesting an oligofluorene (double bonds) can show change in conformation over two order of magnitude faster than those assisted by main vibrational relaxation mode. The molecular picture commented by Mukamel [7] indicates that laser light can bring fluorene into flat geometry much more quickly that was previously thought. The picture described by us is actually an abstract one yet it speaks out the phenomenon:

After the first pulse the system evolves to get at a q-state say $\mathbf{C}(t_1)$; the experiment continues with a second delayed pulse with initial time state $\mathbf{C}(t_1)$; at this point some of double primed base states are not yet accessible. Amplitudes at $^1\Sigma_{j'}$, $^1\Pi_{j'}$ and $^3\Delta_{j'}$ base states may be different from zero; the new pulse may find these as root states. The second photon pulse can put non-zero amplitudes at base states $^3\Delta_{j''}$, $^1\Pi_{j''}$ and $^1\Sigma_{j''}$, e.g. eq.(11). The $^3\Delta_{j''}$ state may last even longer to originate e.g. luminesce as the case may be (e.g. more complex systems).

Any materiality providing a set of base states as those described above will show quantum torsional dynamics; and this independent from the size. Such is the advantage of a Hilbert/Fock space presentation as given here.

### 6.2. From quantum evolution

The question explored with the present quantum physical model shows classical mechanical pictures do not find a natural place yet this is not to say a full elimination would follow. Reify a quantum state is not granted; see Mermin for a cogent discussion in ref. [14], see also [9] but sober use of such models might help. Thus, even if quantum states are sustained by the elementary material elements and do not represent that materiality in a classical mechanical sense as objects the picture can be enriched by sensing presence via the time dependent linear superpositions.

Of course, there is a sort of classical limit if one retains the orthodox statistical interpretation [11]. Here, each element of the ensemble is in a particular base state: all amplitudes zero but one. With this view a "picture" of the transition structure would emerge. Albeit in such scheme all quantum dynamics possibilities are lost.

Furthermore, there is a manner to hint at changes the materiality can be submitted. The excited electro-nuclear states appear as photonic ones where the stitching frequency relates to root states. This one may be a standard ground state, e.g. "vertical" excitation. But, the root may also be another photonic state thereby conveying a second (or third, etc.) photon label into the picture; the case examined here shares this character.

For the linear coherent states covering different j-levels the materiality sustains a number of base states so that the concept of object fades away. Instead one can see all possibilities for change; the question about what "structure" is to be shown by the materiality is not compulsory. Yet, this quantum state operates as q-transition state (q-TS).

The key issue is sensing what response such q-TS may show towards particular probes. By targeting this state for response on the amplitudes related to the j''-domain (E-isomer) one can get a picture of time evolution similar to the one gotten by Zewail in his pump-probe experiments [15]. Yet now classical vibration motion is not required; for many people this picture might well do if we only use this *concept as tool*.

We must realize that within photon entangled sector, resonance with direct product states alters the bare states. Such changes would enforce new response properties. The physical base states highlight this state of affairs.

Of course, molecular structure matters. A way to get at it: the coherent state must undergo a decoherence process for example the one indicated with symbol $-\Rightarrow$. Below another situation is examined. A physical quantum state before such "catastrophe" endows the materiality with new response possibilities (properties).

**7. Discussion**

A photonic view designed to help understanding the quantum physical world is a key element of this approach. The picture of objects occupying energy levels is out and abstract physical quantum states are in. Yet taking concepts as tools rather than pictures would be the attitude to be taken. These conceptual quantum states are sustained by elementary material elements, yet no hint obtains as to their "motions" in real space (see also [1,12,13]). The selection here of the name photonic(s) includes the photon and material base states in a unified concept; thus, $|1_\omega\rangle$ stands for a basis state that reckons one EM energy quantum that might be available in the laboratory field at frequency $\omega$; yet the symbol does not represent a materiality as such, it is only a label. If there is no one photon accessible, the amplitude affecting such base state in the corresponding quantum state is zero. One does not erase base vectors. Only the amplitudes can be modulated and eventually be put equal to zero.

In this paper the emphasis is put on constructing basis sets. To generate time evolution the Hamiltonian must now include coupling effects, namely, extra EM sources, spin-orbit couplings and combinations thereof. The secular equations would provide means to simulate a large range of situations in a rigorous quantum physical framework; see also S. Mukamel in ref.[16] to find different ways.

Once all the above factors are incorporated the q-path to be followed is schematized by eq.(11). These sequences stand for a torsional process albeit now cast in a fundamental quantum dynamic framework. These quantum dynamical effects would appear in the time evolution via **C**(t), which is the surrogate of a wave function once a base set is made explicit; e.g. eq.(1) and base set eq.(9).

The role played by consecutive pulses in the experimental work appeared more explicitly. One-photon processes are not effective in activating the femtosecond torsional change.

Of special interest is the charting that can be made to experimental conditions. Spectral responses are at the foundation of this approach. The quantum numbers can tell us the story. Similar methodology is in current use in the experimental world [17].

At the origin of all this resides a different view about quantum physical processes. The concept of object has receded and leaves the scene to elements of Hilbert/Fock space. The advantage is that if necessary one can reconstruct a structural chemistry in a pictorial manner at a semi-classic level. Thus concepts are used as tools rather than pictures. This is the motto.

Quantum measurements only probe quantum states that are responses modulated by the amplitudes [1]. Amplitude different from zero signals a possible active root state. Thus, an appropriate multi-frequency probe would hence produce responses characteristic of given quantum states [1,12]; recording the response in intensity regime the relative intensities would relate to the square modulus of contributing terms [1,13]. The coherent time evolution is stopped, albeit standard Copenhagen interpretation becomes clearly inadequate; notwithstanding some of their concepts taken as tools can still render service.

If we go back at the level of theory described above, symbols like $+\Rightarrow$ and $-\Rightarrow$ do not show up as processes but as *events*. Predictability of such events within quantum theory is not feasible. Thus, time reversibility is spoiled for quantum time evolution taken as a laboratory process; time reversal has limited scope in laboratory (real) world.

Observe that a conformational change does not necessarily need pictures (cartoons) to understand foundational aspects of the phenomenon. The key lies in identifying those EN states that are significant together with the quantized photon fields. Moreover, it is Hilbert/Fock space that permits setting up conditions for coherence. In this manner, even very large systems such as light- harvesting antenna complexes can show coherent behavior [18]. To isolate a "product" decoherence plays key roles.

The present approach by assigning a tool-nature to concepts generalizes them so that can be developed for atomic/molecular systems and makes them useful for Nano- as well as macroscopic systems. But it is the concept of physical quantum state that is the novelty that we use now in a new dimension.

So far we were used to representational characters of the theoretical elements. Yet here, what is required is the presence of the basic material elements that would sustain the quantum states; which-way descriptions do not make sense now. As long as the quantum system evolves in the coherent domain any classical picture is not warranted. Nevertheless, (semi-) macroscopic quantum objects acting as scattering centers, more or less complex, permit engineering probing/ measuring devices. Real space distributions of material elements sustaining entanglement subdomains lead to quantum detectors or emitters. Quantum phenomena are sustained by any materiality of the kind examined here; it is not a direct function of mass. These q-objects are not exclusively microscopic in nature.

The problem for us is that the event associated to say spontaneous emission or absorption cannot be described in terms adequate to Hilbert and Fock spaces that are not commensurate to real space. Nothing of interest is gained in pronouncing principles that make acceptable what it is not. Duality principles were most useful during the early days of quantum mechanics. Today, only weirdness is gained. And it is not quantum mechanics but our stubborn use of self-contradictory principles based on classical physics that produces a lack of understanding. This situation actually calls for thinking of the foundations of quantum science. And it is pointing to a change of epoch that will require new concept-tools and not pictures from the past to gain new paradigmatic grounds.

**Acknowledgments**

The author is most grateful to all his coworkers referenced here. R. Contreras, University of Chile (Science Faculty) is warmly thanked for inviting the present author for discussions and a lively Seminar.
Part of this work was presented also in the Seminar of Prof. A. Toro Labe from Catholic University (Chile); the author acknowledges lively discussions. Prof. J. J.M. Ugalde, Euskal Herriko Unibersitatea Donostia, Euskadi, Spain and his group are warmly thanked for enlightening discussions.

**References**

1. O. Tapia, Adv. Quantum Chem. **61**, 49 (2011).
2. G.A. Arteca and O. Tapia, Phys. Rev. A **83**, 032311 (2011).
3. G.A. Arteca, J.M. Aulló, and O. Tapia, J. Math. Chem. **50**, 949 (2012).
4. O. Tapia, Adv. Quantum Chem. **56**, 31 (2009).
5. O.Tapia and G.A. Arteca, Internet Electron. J. Mol. Design **2**, 454 (2003). [/www.biochempress.com/cv02_i07.html]
6. J. Clark, T. Nelson, S. Tretiak, G. Cirmi and G. Lanzani, Nature Phys. **8**, 225 (2012).
7. S. Mukamel, Nature Phys. **8**, 179 (2012).
8. D. Roca-Sanjuán, F. Aquilante and R. Lindh, WIREs Comput.Mol.Sci. **2**, 585 (2012)
9. R.Crespo, M.C. Piqueras, J.M. Aulló, O. Tapia, Int. J. Quantum Chem. **111**, 263 (2011)
10. C. Cohen-Tannoudji, J. Dupont-Roc, and G. Grynberg, *Photons and Atoms*: *Introduction to Quantum Electrodynamics* (John Wiley &Sons, New York, 1997).
11. E. Ballentine, *Quantum Mechanics: A Modern Development* (World Scientific, Singapore, 1998).
12. O. Tapia, *Quantum Physical Chemistry: Quantum physics primer* (Publicacions Universitat Jaume I, Castelló de la Plana, Spain, 2012) Amazon.es; ISBN: 978-84-8021-828-3
13. H. Fidder and O. Tapia, Int.J.Quantum Chem. **97**, 670 (2004)
14. N.D. Mermin, Phys.Today, May pp.8 (2009)
15. A. H. Zewail, FEMTOCHEMISTRY, vol. I & II, World Scientific, Singapore (1994)
16. S. Mukamel, *Principles of Nonlinear Optical Spectroscopy* (Oxford University Press, 1995)
17. D.A. Pantazis, W. Ames, N. Cox, W. Lubitz and F. Neese, Angew.Chem.Int.Ed. **51**, 9935 (2012)
18. T.R. Calhoun *et al*. J.Phys.Chem.B **113**, 16291 (2009)